\title{Challenges and pitfalls of partitioning blockchains}
\author{
        Enrique Fynn \\
	Universit\`{a} della Svizzera italiana (USI)\\
	Switzerland\\
        \and
        Fernando Pedone\\
	Universit\`{a} della Svizzera italiana (USI)\\
	Switzerland\\
}
\date{\today}

\documentclass[10pt,conference]{IEEEtran}
\usepackage{subcaption}
\usepackage{listings}
\usepackage{hyperref}
\usepackage{todonotes}
\usepackage{graphicx}
\usepackage{algorithm}
\usepackage{varwidth}
\usepackage[noend]{algpseudocode}

\let\oldReturn\Return
\renewcommand{\Return}{\State\oldReturn}

\begin{document}
\maketitle
\begin{abstract}

Blockchain has received much attention in recent years.
This immense popularity has raised a number of concerns, scalability of blockchain systems being a common one.
In this paper, we seek to understand how Ethereum, a well-established blockchain system, would respond to sharding.
Sharding is a prevalent technique to increase the scalability of distributed systems.
To understand how sharding would affect Ethereum, we model Ethereum blockchain as a graph and evaluate five methods to partition the graph.
We assess methods using three metrics: the balance among shards, the number of transactions that would involve multiple shards, and the amount of data that would be  relocated across shards upon repartitioning of the graph.
\end{abstract}

\section{Introduction}

Blockchain technology has gained much traction since Satoshi Nakamoto introduced Bitcoin in 2008~\cite{bitcoin_nakamoto}.
Several blockchain systems have been developed after Bitcoin, some containing serious flaws~\cite{blockchain_in_the_wild}.
A blockchain is a distributed, decentralized and secure ledger or, more technically, a geographically replicated state machine that tolerates byzantine failures.
In a blockchain, blocks contain transactions submitted by the users (e.g., transferring currency from one user's account to another user's account).
Blocks are produced by the miners, each block cryptographically linked to the previous one, forming a chain.
To produce a valid block, miners must solve a cryptographic puzzle.
The miner whose valid block makes it to the canonical chain receives the block reward.



%

\begin{figure*}[ht!]
  \centering
  \includegraphics[width=1\textwidth]{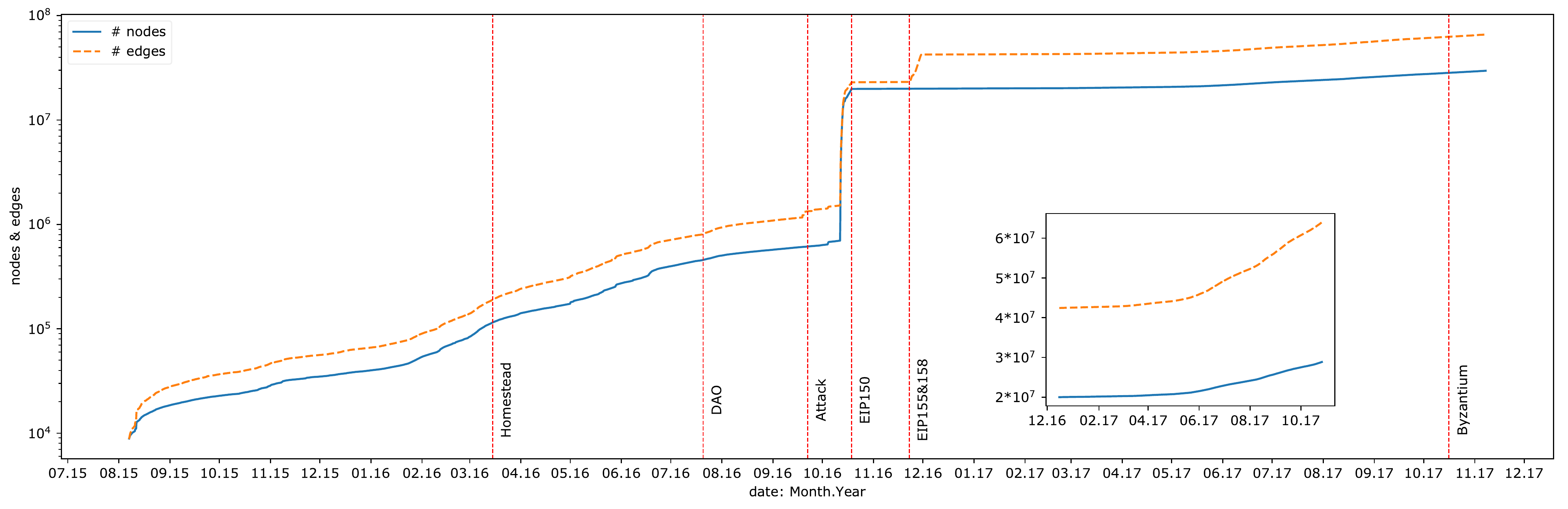}
  \caption{Ethereum graph evolution in number of vertices (accounts and smart contracts) and edges (transactions)}
  \label{fig:ethereum_graph}
\end{figure*}

Even though Bitcoin admits some level of programming, its scripting language is quite restrictive. 
While this decision leads to simple transactions (e.g., possibly with fewer bugs), it hinders the applicability of Bitcoin.
Differently from Bitcoin, Ethereum provides a more generic Turing-complete language to interact with the blockchain.
Consequently, Ethereum has served as a platform for other services and hundreds of applications have been developed based on it.
On the one hand, Ethereum's generality has contributed to its widespread success.
Fig.~\ref{fig:ethereum_graph} shows the evolution of the ``Ethereum blockchain graph," where vertices are accounts and smart contracts (i.e., procedures created and stored in the blockchain and instantiated by user transactions) and edges are interactions between accounts and contracts, resulting from transactions.
Until around October 2016, the number of Ethereum accounts, contracts and transactions grew exponentially over time;
from then, the growth has been superlinear (see zoomed in graph in Fig.~\ref{fig:ethereum_graph}).
On the other hand, Ethereum's success presses its developers to scale the architecture---currently, regular Ethereum members must store and process the whole blockchain.

%


The typical distributed systems recipe for scaling performance is to shard (or partition) the state of the application.
If the partitioning is such that most application requests can be executed within a single shard and the load among shards is balanced, then performance scales with the number of shards.
Unfortunately, few applications can be optimally partitioned (i.e., all requests fall within a single shard and load is balanced among shards) and so the system must handle requests that span multiple shards.
There are two classes of solutions when it comes to handling a multi-shard request:
(a)~having the involved shards coordinate and execute the request in a distributed fashion (e.g., Spanner~\cite{spanner}, S-SMR~\cite{SSMR}) and
(b)~moving the necessary state to one shard that will execute the request locally (e.g., \cite{DSSMR}).
In either case, one pitfall of sharding is that if the application state is poorly partitioned, overall system
performance will most likely decrease, instead of increase, due to the overhead of multi-shard requests.

%
%

In this paper, we consider the problem of partitioning Ethereum's blockchain graph.
The graph is generated by analyzing calls within Ethereum accounts and contracts.
Accounts and contracts can call each other in specific ways in a transaction, and a transaction can lead to multiple calls to different accounts and contracts. 
When a call happens, an edge in the graph is created between caller and callee.
Additionally, we assign weights to vertices and edges to account for their frequency in the blockchain.

The general question we seek to answer in this study is: 
How would Ethereum blockchain graph respond to sharding?
To answer the question, we use five different techniques to partition different instants of the graph over time, from its conception in August 2015 to the end of 2017.
For each technique, we consider the balance among shards, the ratio of edge-cuts versus total edges (an edge-cut represents a multi-shard request), and the amount of data that would be relocated across shards upon a repartitioning of the system.

It is not our goal to propose mechanisms for Ethereum to handle multi-shard transactions.
This aspect is orthogonal to our study and has been a hot topic of discussion in Ethereum's mailing lists.
Our study, however, sheds some light on what one can expect from sharding Ethereum.

Finally, we have made the data set used in the experiments, built from Ethereum blockchain, publicly available in easily understandable format.\footnote{\url{https://dslab.inf.usi.ch/ethereum_trace/}}
The data set can be used for further analysis and benchmarking.

\section{Partitioning Ethereum}

In this section, we briefly describe Ethereum (\S\ref{sec:background}), explain how we build Ethereum's blockchain graph (\S\ref{sec:ebg}), and detail the partitioning methods used in the study (\S\ref{sec:part}).

\subsection{Background}
\label{sec:background}

Ethereum was the first general-purpose blockchain conceived~\cite{ethereum_gavin}. 
Users interact with Ethereum's blockchain by sending a transaction from a user account.
The transaction may transfer some \emph{ether}, Ethereum's underlying currency, to another account or activate a \emph{smart contract}.
When transferring currency from an account to another, Ethereum works similarly to Bitcoin.
The ability to execute arbitrary code in the form of contracts grants Ethereum much more flexibility than Bitcoin.

A smart contract can read and modify internal storage (i.e., a database mapping 32-byte keys to 32-byte values), transfer currency to one or more accounts, and trigger the execution of other contracts.
Contracts are written in Ethereum-specific binary format and executed by the Ethereum Virtual Machine (EVM), a 256-bit stack-based virtual machine. 
Contracts can be created and destroyed by other contracts or accounts.
After the contract is created, it resides in the blockchain.
The initial contract state can be set by using an initialization code that is only executed in the contract's creation. 

At the beginning of a transaction, users have to define the maximum ether they are willing to pay for the execution of their transaction (specified in \emph{gas} and \emph{gas price}). 
Determining the actual cost of a transaction is not obvious.
In fact, there is no way to tell whether a contract will end or not, since this problem is reducible to the halting problem, which is undecidable.
Users can estimate the cost of a transaction from the transaction's instructions and the cost of each instruction.

Miners include transactions in a block based on their estimates of the transaction cost and the amount the user is willing to pay for the transaction.
When the block with the transaction is included in the blockchain and the transaction is executed, the miner receives the ether the transaction consumes.
The scheme is slightly more complex, for example, to cover cases in which transactions fail or run out of gas, but these aspects are not important in the context of our study.

Ethereum has experienced a rapid growth since its inception.
Moreover, Ethereum's consensus rules have been revised (i.e., \emph{forked}) many times.
In Fig.~\ref{fig:ethereum_graph}, we can see the system growth in number of vertices (accounts and smart contracts) and edges (transactions).
Each vertical dashed line in Fig.~\ref{fig:ethereum_graph} shows a fork of the system and the infamous attack that exploited Ethereum's vulnerabilities. 
During the attack period, the number of vertices and edges increased by one order of magnitude.
Afterwards the system continued to grow quickly, although at a lower pace than prior to the attack.

\subsection{Ethereum's blockchain graph}
\label{sec:ebg}

We build a directed graph based on the transactions in Ethereum's blockchain. 
In the graph, vertices represent accounts or contracts, and edges represent interactions involving accounts and contracts.
All the following cases lead to an edge $e$ from vertex $v_1$ to vertex $v_2$ in the graph:
a user with account $v_1$ submits a transaction that transfers some currency to account $v_2$ or that activates contract $v_2$;
contract $v_1$ transfers ether to account $v_2$ or activates contract $v_2$.

\begin{figure}[h]
  \centering
  \includegraphics[width=0.50\textwidth]{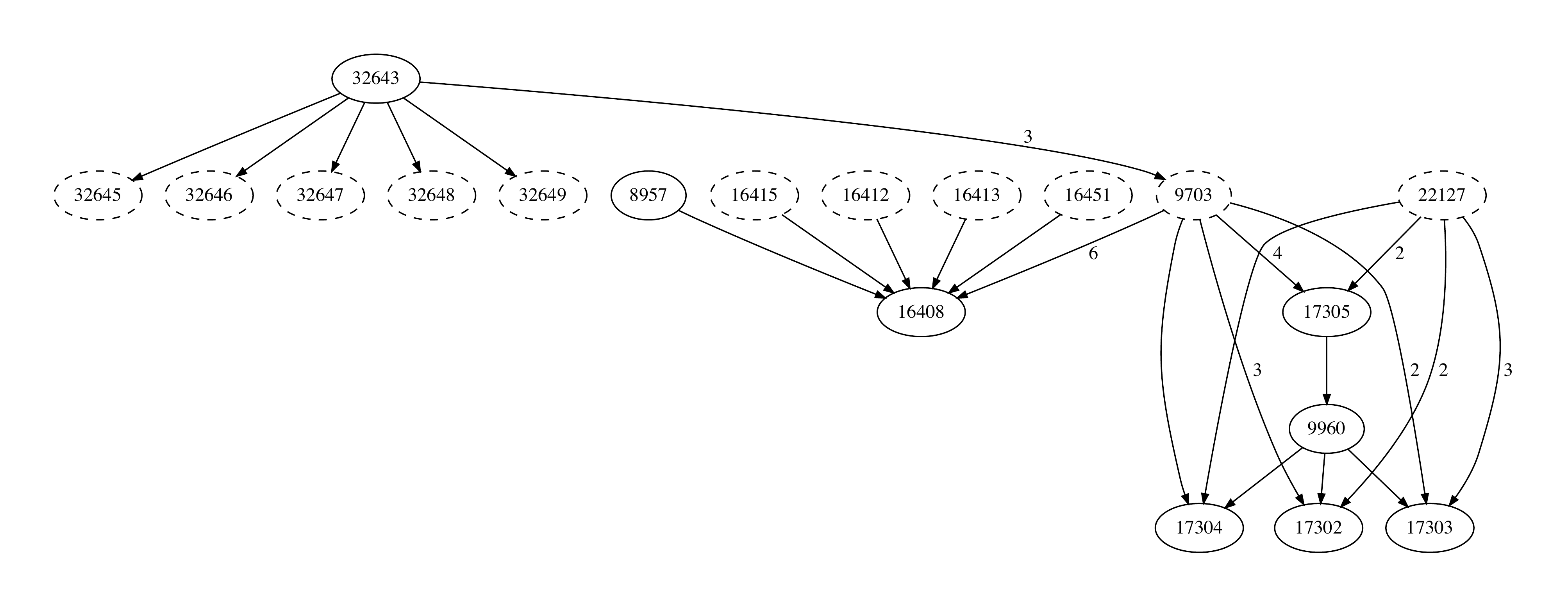}
  \caption{Subgraph with  accounts (full line nodes), contracts (dashed line nodes), and their dependencies (arrows).}
  \label{fig:graph_example}
\end{figure}

\begin{figure*}
  \centering
  \begin{subfigure}[b]{1\linewidth}
  \includegraphics[width=\textwidth]{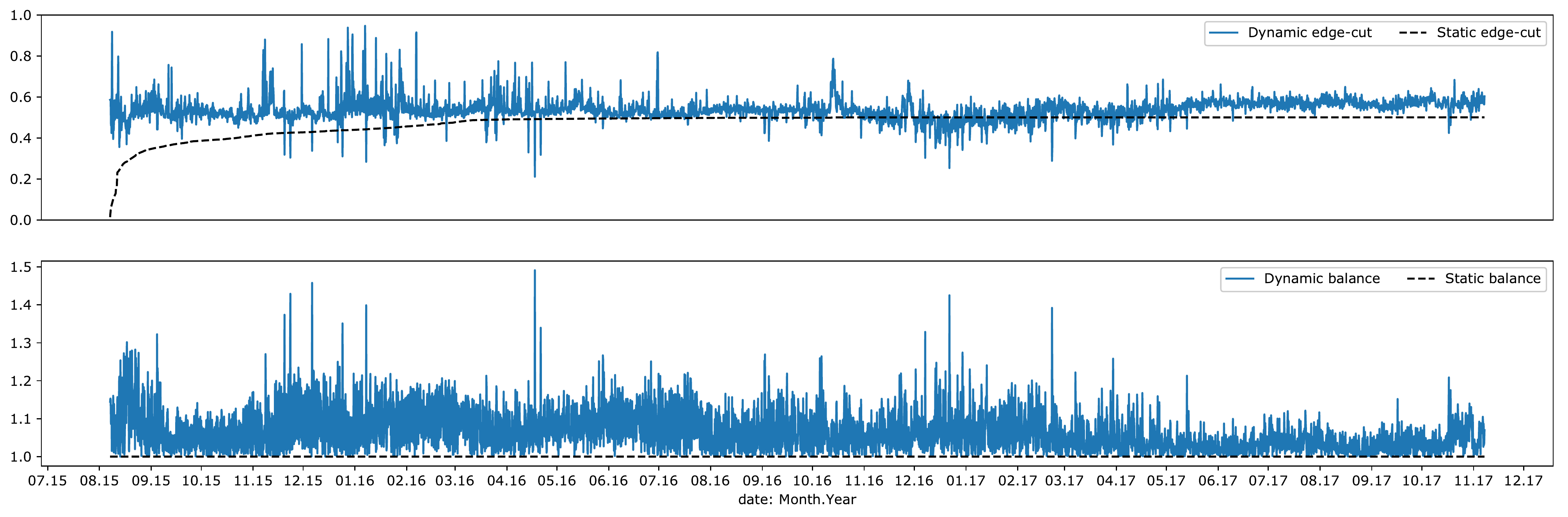}
  \caption{Hashing}
  \end{subfigure}
  \begin{subfigure}[b]{1\textwidth}
  \centering
  \includegraphics[width=\textwidth]{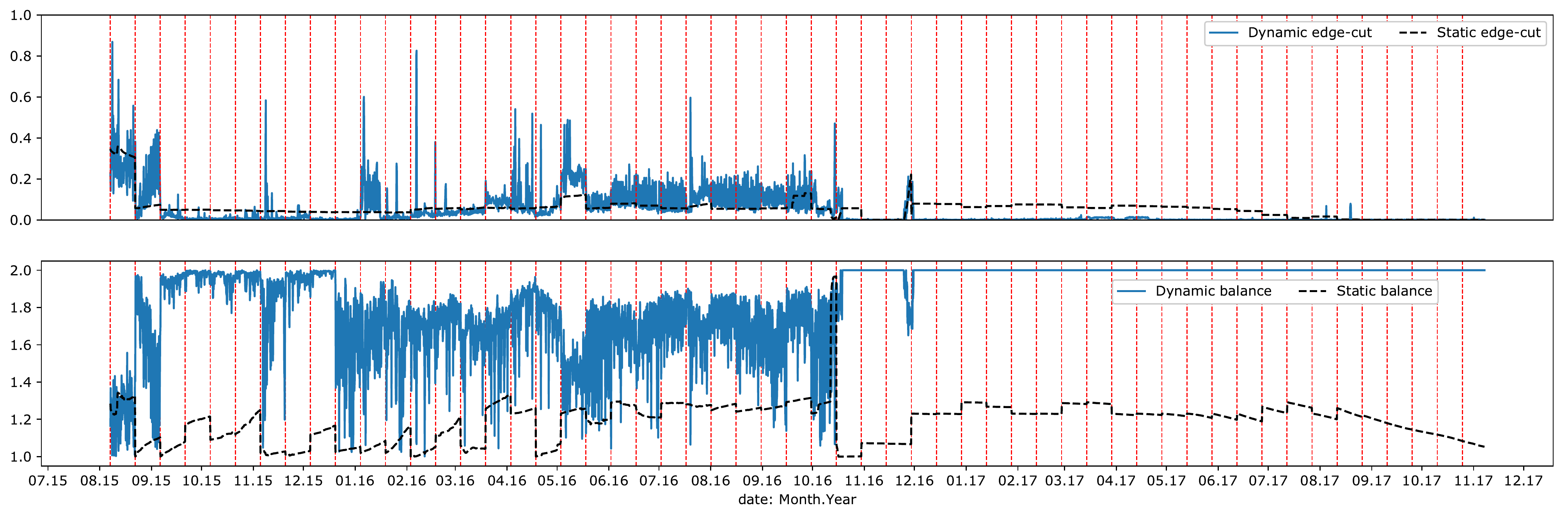}
  \caption{METIS}
  \end{subfigure}
  \caption{The performance of (a) hashing and (b) METIS when Ethereum is partitioned in two shards.}
  \label{fig:metis_partitioning_2p}
\end{figure*}

\pagebreak
Fig.~\ref{fig:graph_example} shows a subgraph of the Ethereum graph in September 2015. 
Full-line nodes represent accounts and dashed-line nodes represent contracts. 
Edges originating in accounts are single transactions, edges originating in contracts can perform different
calls in the same transaction, both to accounts and other contracts. 
The weight in each edge denotes the number of times the interaction happened; when no weight is specified, the interaction happened once.
For example, contract 9703 was instantiated 18 times in the subgraph, 13 times in transactions starting at 8900, 3 times in transactions starting at 8930, and 2 times in transactions starting at 17303.
As part of its execution, contract 9703 transferred ether twice to accounts 9960, 17257 and 17265.
Notice that in the complete graph, there is no contract without at least one incoming edge.


\subsection{Partitioning methods}
\label{sec:part}

We formulate the problem of graph partitioning as follows: 
Given a graph $G = (V, E)$ and a number of partitions $k$, a partition $p_i$ with $1 \leq i \leq k$ is a subset of $V$ such that $\bigcup p_i = V$ and $\bigcap p_i = \emptyset$.
That is, the graph is partitioned in $k$ partitions and partitions are disjunct.

The problem of balanced graph partitioning is NP-Complete~\cite{NPC} and despite being studied for long (e.g., VLSI designs~\cite{VLSI}), there are several ways to define what constitutes a good partitioning.
Intuitively, a good partitioning is defined as one that minimizes the edges connecting two partitions (edge-cut) while 
maintaining each partition balanced.

More formally, let $C(p_i)$ be the set of edges that connect vertices in $p_i$ with vertices in $V \setminus p_i$. 


\begin{equation}
	edge\mbox{-}cut=\frac{\sum_{i=1}^{k}|C(p_i)|}{|E|}
	\label{eq:edgecut}
\end{equation}

\begin{equation}
	balance=\frac{max_{1 \leq i \leq k}(|p_i|) \times k}{|V|}
	\label{eq:balance}
\end{equation}

From Eq.~\ref{eq:edgecut} we get the edge-cut percentage and from Eq.~\ref{eq:balance}, the relation
between the most unbalanced partition and the others. For example, if $k=2$ and $edge\mbox{-}cut$ and $balance$ are, 
respectively, 0.2 and 1.3, then 20\% of the edges are across partitions and one partition has 30\% more vertices than the average.
Ideally, $edge\mbox{-}cut$ is 0 and $balance$ is 1.


With Eqs. \ref{eq:edgecut} and \ref{eq:balance}, only static aspects of the graph are taken into account. 
We can enrich the graph by assigning weights to vertices and edges to capture the frequency that accounts, contracts, and their interactions appear in the blockchain.
By assigning weights to the edges, we can try to avoid cutting frequently used edges; by assigning weights to the vertices, we can better balance the load in the system.
We refer to Eqs. \ref{eq:edgecut} and \ref{eq:balance} of a weighted graph as dynamic edge cut and dynamic balance, respectively. 
The dynamic edge cut and dynamic balance give us a more accurate view of the system's executed cross-shard transactions and load.

In the following we describe the five partitioning methods we used to partition Ethereum blockchain graph.


\begin{itemize}
\item \emph{Hashing.}
A straightforward way to partition the graph is to hash the vertex unique identifier and use the result (modulo the total number of shards $k$) to determine the shard the vertex belongs to.
This is a common scheme to shard data.
Moreover, if the hash distribution is uniform, we can hope to obtain an optimum static balance among shards.
The edge cut (number of multi-shard transactions), however, tends to increase with the number of shards.
For instance, when $k=8$ in our experiments, multi-shard transactions account for 88\% of the total,
despite the vertices being perfectly distributed.

\item \emph{Kernighan-Lin algorithm \cite{KL}.}
In this method, we periodically partition the system based on the transactions executed in the period.
The system starts in some partitioned state (e.g., computed using hashing).
Periodically, based on the transactions executed in the period, each shard identifies vertices that if moved to other shards would minimize edge-cuts.
Each shard sends to an oracle the selected vertices and with the information from all shards the oracle computes a $k \times k$ probability matrix. 
The oracle calculates the probability that each shard should move its selected vertices to the other shards so that at the end shards remain balanced.
The oracle then sends the matrix to all the shards, which exchange vertices with each other based on the probability matrix.
Intuitively, the algorithm tries to reduce dynamic edge-cuts while keeping the shards dynamically balanced.
This approach has been used to partition large graphs \cite{facebook_gp}.
%
%
%

\item \emph{METIS.} \label{sec:partial_metis}
In this method, we periodically partition the graph with METIS~\cite{METIS}. 
To do so, we input METIS with the current graph (i.e., up to the moment of partitioning) and shard vertices based on METIS output.
METIS strives to minimize edge-cuts while keeping shards balanced.
We aim to reduce dynamic edge-cuts by assigning weights to the edges of the graph.
When an account (or contract) appears for the first time (or is created), it has to be assigned to some shard.
This is done by inspecting all the accounts involved in the transaction and picking the shard that minimizes edge-cuts; if more than one exists, we maximize the balance.
With this approach, every time the graph is partitioned, METIS can move vertices back and forth between shards, since it is not part of METIS objectives to minimize the number or vertices that change shard between successive executions of the algorithm.

\item \emph{R-METIS.}
While using the technique above with large graphs, we realized that METIS sometimes does not produce good results.
We therefore revisit the previous technique by using as input for METIS a reduced graph.
This graph contains all accounts, contracts, and their interactions within a fixed window of time (two weeks), which starts at the last (re)partitioning.

\item \emph{TR-METIS.}
This is essentially the method above where instead of triggering a repartition at constant time intervals, we set a threshold on the dynamic edge-cut and dynamic balance.
When the threshold is reached, we run METIS to compute a new partitioning, which will hopefully reduce the number of transactions across shards.
The motivation for the technique is to reduce unnecessary repartitioning.

\end{itemize}

\section{Results}

We assessed the five partitioning methods described in the previous section in configurations with 2, 4 and 8 shards.
In each case, we computed the following metrics: static and dynamic edge-cut and balance and the number of vertices that change shard after the graph is repartitioned (i.e., number of moves).

Figure \ref{fig:metis_partitioning_2p} shows the results for hashing and METIS with two shards.
Each data point corresponds to a four-hour window.
Repartitioning takes place every two weeks in case of a periodic repartitioning, marked by vertical dashed lines in Figure \ref{fig:metis_partitioning_2p}(b).
Hashing provides optimum static balance since each shard is assigned an equal number of vertices.
This does not mean, however, that the load experienced by the shards is the same (see dynamic balance).
With respect to static edge-cuts, with two shards hashing leads to about 50\% of transactions across shards.
METIS provides a much lower edge-cut, both static and dynamic, at the expense of dynamic imbalance among shards.
Notice that dynamic balance is near two.
This happens because vertices in one shard are much more ``active" than vertices in the other shard, even though both shards contain similar number of vertices.
The effect is particularly visible after the September 2016 attack, when a large number of dummy accounts were created.

\begin{figure*}
  \centering
  \begin{subfigure}[b]{1\textwidth}
     \includegraphics[width=1\linewidth]{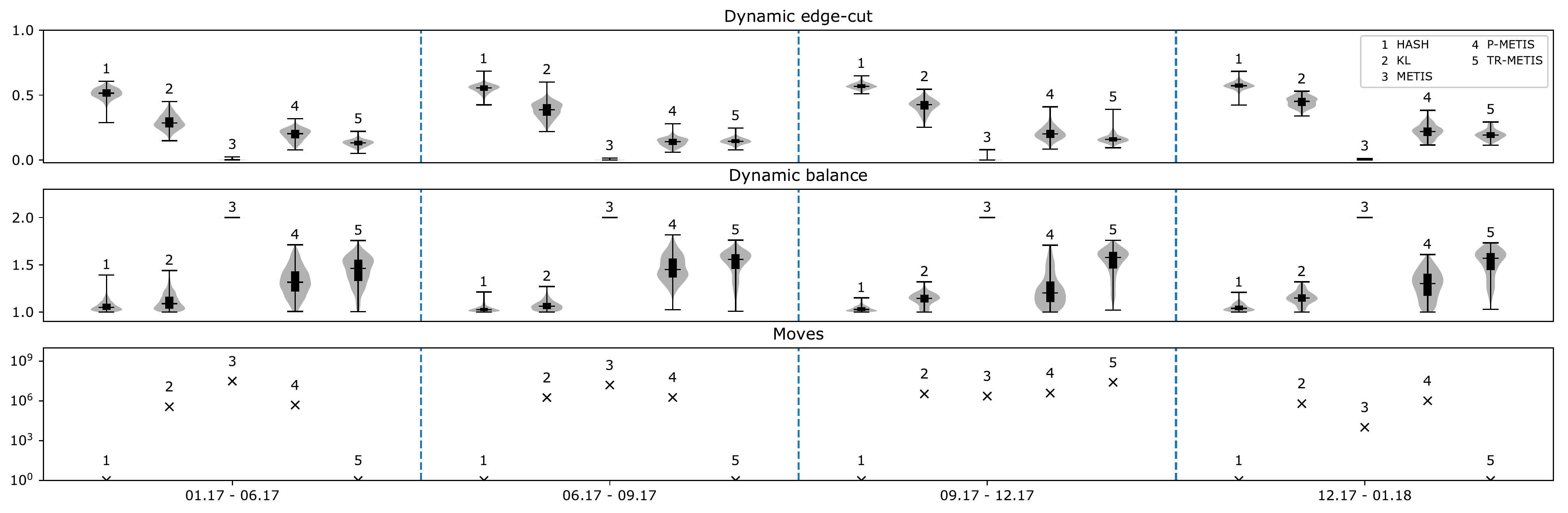}
     \caption{Configurations with 2 shards}
  \end{subfigure}
  \begin{subfigure}[b]{1\textwidth}
     \includegraphics[width=1\linewidth]{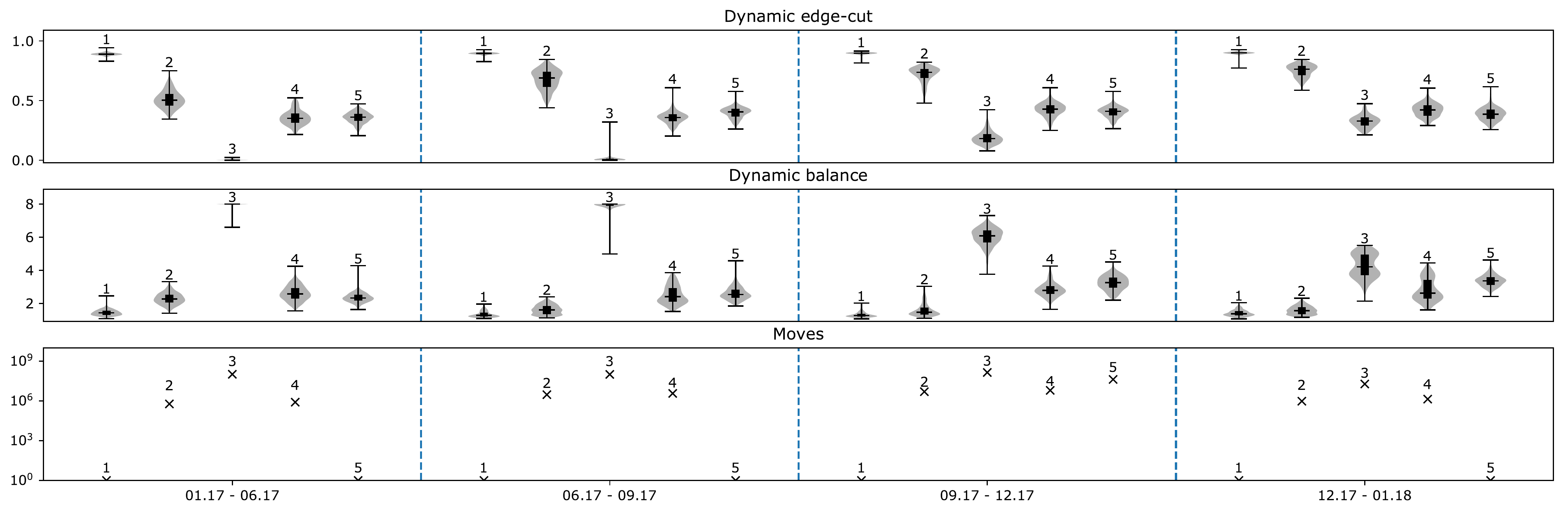}
     \caption{Configurations with 8 shards}
  \end{subfigure}
  \caption{Box-and-whisker with violin plot (density plot) for five partitioning methods using Ethereum transactions in 2017 (whiskers show minimum and maximum values, bottom and top of the box show first and third quartiles, and the band inside the box shows the median).}
     \label{fig:Ng1} 
\end{figure*}

\begin{figure*}[h]
  \centering
  \includegraphics[width=\textwidth]{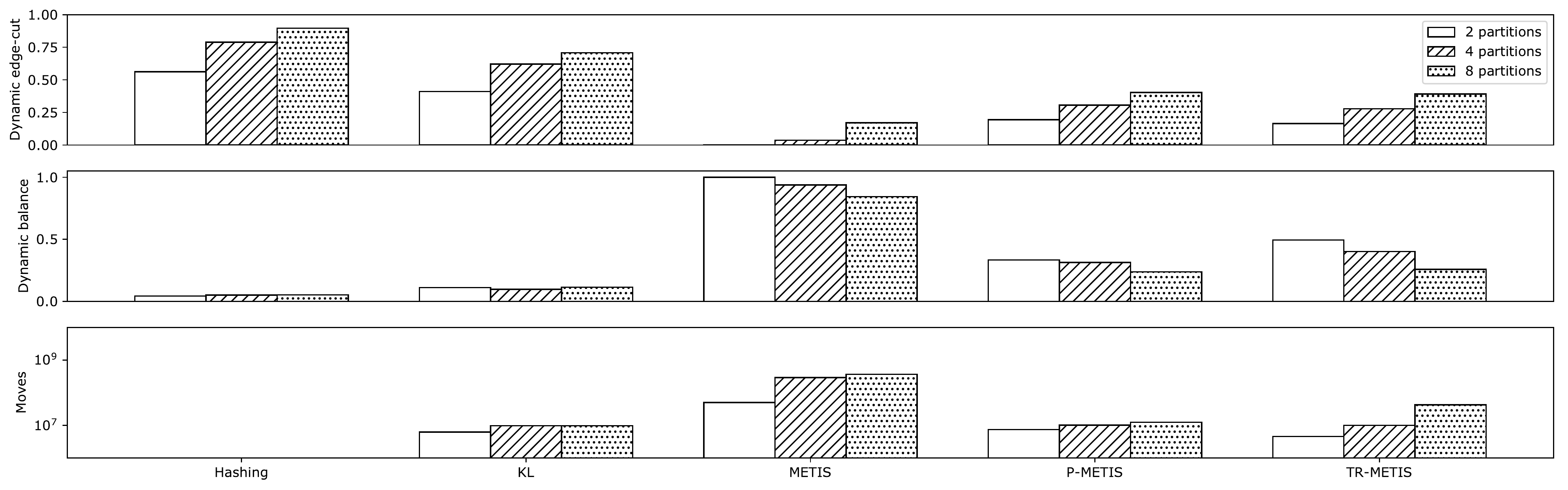}
  \caption{Dynamic edge-cut, dynamic balance and total number of moves of various techniques with increasing number of shards.}
  \label{fig:bar_all_techniques}
\end{figure*}

For brevity, we do not present detailed results for the other techniques and configurations with 4 and 8 shards.
This data is available online (see footnote 1).
Instead, in the following we summarize our findings.

Fig.~\ref{fig:Ng1} shows results for the five partition methods in configurations with 2 and 8 shards, using transactions executed in 2017, which represents the bulk of transactions in Ethereum.
We show dynamic edge-cut, dynamic balance and total number of moves in the period.

We draw the following conclusions from these results.
\begin{itemize}
\item There is a clear compromise between edge-cut and balance, and no technique clearly stands out.
This suggests that \emph{based on current workload, it is unlikely that Ethereum can be partitioned in such a way as to produce both low edge-cut and good balance among shards}.

\item Within a configuration (i.e., 2 or 8 shards), the behavior of the various partition methods does not change substantially over time, although it does change from one configuration to the other.
We take a closer look at this aspect at the end of this section.

\item Hashing leads to a fair dynamic balance among shards, albeit at the cost of a large number of dynamic edge-cuts.
Interestingly, hashing does not outperform KL with respect to dynamic balance, but performs consistently worse than all other techniques for dynamic edge-cuts.
There are no moves since partitioning depends on vertex id only and once assigned to a shard a vertex remains in the assigned shard.

\item KL reduces dynamic edge-cuts while maintaining shards balanced. 
The various iterations of the technique lead to a large number of vertices changing shards.
One pitfall of this method is that partitioning optimizes for a local minima.

\item METIS trades balance for edge-cuts.
However, as previously discussed, this is mostly an anomaly that resulted from the October 2016 attack, when a large number of dummy accounts were created and never used again.
Although METIS statically balances the graph, one shard contains most meaningful accounts and transactions execute in a single shard.

\item R-METIS considers just the subgraph formed after a repartitioning.
This helps ignore vertices created and never used again. 
Vertices only used once create an artificial balance among shards. With this technique
we managed to get a lower dynamic balance. 
This technique resulted in a more dynamically balanced system.

\item TR-METIS improves on the previous technique by reducing the number of vertices moved across shards.
This essentially happens because we trigger a repartitioning based on edge-cut and balance values. 
We adjust thresholds to trigger a repartitioning in such a way that the performance does not diverge much from the previous technique.
The result is a dramatic decrease in the number of moved vertices, without compromising edge-cuts and balance, when compared to R-METIS.
\end{itemize}

Note that reducing the number of vertices that change shards after a repartitioning of the system is important for performance.
If we were to move one vertex from one shard to another, we ought to move the entire state of the vertex.
If the vertex is a contract, that would result in moving the entire contract storage to another shard.

Fig~\ref{fig:bar_all_techniques} compares partitioning techniques with respect to dynamic edge-cut and balance and total number of moved vertices while varying the number of shards.
In these executions we used data from the beginning of Ethereum up to the end of 2017.
In order to show balance for different configurations in the same graph we normalized the results with the number of shards (i.e., $\mathit{balance\ value-1} / \mathit{number\ of\ shards-1}$).
As before, low values mean better balanced shards.

In all techniques, dynamic edge-cut becomes worse as the number of shards increases (top graph), although METIS-based techniques outperform hashing and KL.
With respect to dynamic balance, Hashing and KL perform better than techniques based on METIS.
Hashing and METIS, however, take extreme ends in the balance versus edge-cut tradeoff.
The number of moves is large in the METIS algorithm, since the partitioner algorithm does not optimize for this aspect.
P-METIS and TR-METIS techniques perform substantially fewer moves because they use a smaller graph.

\section{Final remarks}\label{future_work}

This study sheds some light on what one could expect from sharding Ethereum.
The results show that even with fairly sophisticated partitioning methods, there is a clear tradeoff between edge-cuts and balance.
This is important since scalable performance  requires  low edge-cut and balanced partitioning.

The results come with some caveats.
First, we assess Ethereum using the real workload, which was not created for a sharded system.
It is possible that if Ethereum is ever extended with the ability to handle sharding then applications will be designed in a different way.
If sharding is made visible to developers, then multi-shard operations could be sometimes avoided, at the expense of more complex applications.

Second, when sharding a blockchain, multiple incentives have to be taken into account.
In case of a generic framework such as
Ethereum, there are three main components that need to be addressed: computation, storage and bandwidth~\cite{crypto_fees}.
All of these components play an important role in partitioning.
For instance, moving state indiscriminately will have both an impact in the bandwidth and storage of the system. 
Designing the correct incentives is crucial to a good partitioning scheme.

\section*{Acknowledgements}

This work was supported in part by the Swiss National Science Foundation under grant number 166132.
%


\bibliographystyle{IEEEabrv}
\bibliography{references}

\end{document}